# The question of charge and of mass.

Voicu Dolocan

*Faculty of Physics, University of Bucharest, Bucharest, Romania*

**Abstract.** There are two long –range forces in the Universe, electromagnetism and gravity. We have found a general expression for the energy of interaction in these cases, $\alpha\hbar c/r$, where $\alpha$ is the fine structure constant and $r$ is the distance between the two particles. In the case of electromagnetic interaction we have $\alpha\hbar c = e^2/4\pi\varepsilon_o$, where $e$ is the gauge charge, which is the elementary electron charge. In the case of the gravitational interaction we have $\alpha\hbar c = GM^2$, where $M = 1.85\times 10^{-9}$ kg is the gauge mass of the particle. This is a giant particle. A system of like charged giant particles, would be a charged superfluid. By spontaneous breaking of a gauge symmetry are generated the Higgs massive bosons. The unitary gauge assure generation of the neutral massive particles. The perturbation from the unitary gauge generates charged massive particles. Also, the Higgs boson decays into charged and neutral particles. The Tesla coil is the user of the excitations of the vacuum.

## 1. What is electric charge, and what is mass?

According to the Standard Model " The electric charge is a fundamental conserved property of certain subatomic particles that determines the electromagnetic interactions". Electrically charged particles are influenced by and create electromagnetic fields. The elementary unit of charge is carried by a single proton and the equivalent negative charge is carried by a single electron. Also, there are up quarks with $(2/3)e$ charge and there are down quarks with $(1/3)e^-$ charge. Two up quarks and a down quark add up to a charge of $1e$ for a proton. One up quark and two down quarks add up to a charge of zero for a neutron[1-3]..

| Particles | Symbol | charge | mass, keV |
|---|---|---|---|
| electron | $e^-$ | $-e$ | 511 |
| positron | $e^+$ | $+e$ | 511 |
| up quark | $u$ | $+(2/3)e$ | 3000 |
| up antiquark | $/u$ | $-(1/3)e$ | 3000 |
| down quark | $d$ | $-(1/3)e$ | 6000 |
| down antiquark | $/d$ | $+(1/3)e$ | 6000 |

Also, there are charged bosons, as mesons, which are composed from a quark and an antiquark.

The quantity of any matter is the measure of it by its density and volume conjointly. This quantity is what we understand by the term mass or body. It was found by pendulum experiments that the mass of a body is proportional to its weight, as was shown by Galilei. Coulomb's law has the same mathematical form as Newton's law of gravity.
There are some questions.
- It is possible that a charged particle with zero rest mass exist?

- What is positive charge and negative charge?
- Since the Higgs bosons give their masses to the elementary particles, are there other bosons which give them their electric charge?
- Why the charge to mass ratios of the various elementary particles are very dissimilar?
- What is the true nature of the electric charge?

Further we will attempt give the answer to some of these quaestions.

## 2. The gauge charge and the gauge mass.

Suppose that two charges $q_1$ and $q_2$ are located at position vectors $\mathbf{r}_1$ and $\mathbf{r}_2$. According to the Coulomb's law, the energy of interaction between these charges is written

$$E_I = \frac{q_1 q_2}{4\pi\varepsilon_o r}$$

where $r = |\mathbf{r}_2 - \mathbf{r}_1|$. The energy of interaction between two electric charges is radial, inverse, and proportional to the product of the charges. Two like charges repel one another, whereas two unlike charges attract. $\varepsilon_o = 8.8542 \times 10^{-12}$ $C^2N^{-1}m^{-2}$ is the permitivity of free space.

In a previous paper [4] we have obtained an expression for the Coulomb's law without taking into account the electric charge. Next we review this problem. In order to write the Hamiltonian for the electron-boson interaction consider a string (rope) on which electron may walk. We have therefore the "tight-rope walking electrons" [5]. The Hamiltonian density for the isotropic elastic continuum is defined by [4-6]

$$H_d = \frac{1}{2\rho}\Pi_n \Pi_n + \frac{DR_l}{2}\frac{\partial u_n}{\partial z_n}\frac{\partial u_l}{\partial z_l} + \frac{\beta R_l}{2}\frac{\partial u_n}{\partial z_l}\frac{\partial u_n}{\partial z_l} - \Psi^+(z)[-\frac{\hbar^2}{2m}\frac{\partial^2}{\partial z^2}]\Psi(z) + \frac{D}{2}s_l\frac{\partial u_n}{\partial z_n}\frac{\partial u_l}{\partial z_l}\Psi^+(z)\Psi(z) + \frac{\beta}{2}s_l\frac{\partial u_n}{\partial z_l}\frac{\partial u_n}{\partial z_l}\Psi^+(z)\Psi(z)$$

(2.1)

We are to sum over repeated indices. The coordonate axes $z_n$ are assumed to be orthogonal. The term in D is the square of the trace of the strain tensor; the term in β is the sum of the squares of the tensor components. $R_l$ is the distance between the two electrons, $s_l$ is the electron displacement near its equilibrium position,

$$\rho = \rho_o + \frac{(D\delta_{nl} + \beta)R_l}{c^2} \qquad (2.2)$$

$\rho_o$ is the massive density of the interacting field if this is a masive field, c is the velocity of the boson waves and $(D\delta_{nl}+\beta)R_l/c^2$ is the density of the interacting field if this is the massless field. $\Pi_n$ are the components of the momentum density, $u_n(z)$ is the displacement of the string (that is of the coupling field) at the position $z$. The Hamiltonian of interaction is

$$H_I = \frac{D}{2}\int\sum_l R_l s_l \frac{\partial u_n}{\partial z_n}\frac{\partial u_l^*}{\partial z_l}\Psi^+(z)\Psi(z)dz + \frac{\beta}{2}\int\sum_l R_l s_l \frac{\partial u_n}{\partial z_l}\frac{\partial u_n^*}{\partial z_l}\Psi^+(z)\Psi(z)dz \quad (2.3)$$

where we have introduced the sum over neighbours, and [4,5]

$$u_l(z) = \frac{1}{\sqrt{NR_l}}\sum_q \left(\frac{\hbar}{2\rho\omega_q}\right)^{1/2}\left(a_q e^{iqz} + a_q^+ e^{-iqz}\right) \quad (2.4)$$

$$s_l = \frac{1}{NR_l}\sum_k \frac{1}{k^2}e^{ik(z-z_1)}\left(c_k + c_{-\kappa}^+\right) \quad (2.5)$$

$\omega_q$ is the classical oscillation frequency

$$\omega_q = \left(\frac{D\delta_{nl}+\beta}{\rho}R_l\right)^{1/2} q \quad (2.6)$$

where $l$ denotes the longitudinal boson. $a_q^+$ and $a_q$ are boson creation and annihilation operators, and

$$\Psi(z) = \frac{1}{\sqrt{NR}}\sum_{k\sigma} c_{k\sigma} e^{ikz}$$

$$\Psi^+(z) = \frac{1}{\sqrt{NR}}\sum_{k\sigma} c_{k\sigma}^+ e^{-ikz} \quad (2.7)$$

where N is the number of the links in the case of the linear lattice, $c_{k\sigma}^+, c_{k\sigma}$ are the electron creation and annihilation operators, **k** is the wave vector of an electron and $\sigma$ is the spin quantum number. Finally, the interaction Hamiltonian may be written

$$H_I = \hbar \sum_{k,k',q,q',\sigma,\sigma'}\{g_k c_k (a_q + a_{-q}^+)(a_{-q'} + a_{q'}^+)c_{k'\sigma'}^+ c_{k\sigma}\Delta(\mathbf{q}-\mathbf{q}'+\mathbf{k}-\mathbf{k}'+\mathbf{k})+$$
$$g_k^* c_k^+ (a_{-q} + a_q^+)(a_{q'} + a_{-q'}^+)c_{k\sigma}^+ c_{k'\sigma'}\Delta(\mathbf{q}'-\mathbf{q}+\mathbf{k}'-\mathbf{k}+\mathbf{k})\} \quad (2.8)$$

where

$$g_\mathbf{k} = \frac{D}{4N^2 R\left(\rho_o + \frac{DR}{c^2}\right)} \frac{\mathbf{qq'}}{k^2 \omega_q} \sum_l e^{ikz_l} \qquad (2.9)$$

where shear term and also the indices *n* were omitted. We use the approximation of nesrest neighbours. In the Kroenecker Δ we use **q' = k, k' = k + q**. In the interaction picture the expectation value of the energy of electron-electron interaction is

$$E_I = 2\sum_{q,k} |g_\mathbf{k}|^2 \frac{1}{(\varepsilon_\mathbf{k} - \varepsilon_{\mathbf{k-q}}) - \omega_q} (n_q + 1) n_\mathbf{k} n_{\mathbf{k-q}} \qquad (2.10)$$

For $n_q = 0$, $n_\mathbf{k}, n_{\mathbf{k-q}} = 1$, N +1, from Eq. (11) one gets

$$E_I = \sum_{k,q} \frac{D^2}{16R^2 k^4} \frac{(\mathbf{qk})^2}{\left(\rho_o + \frac{DR}{c^2}\right)^2 \omega_q^2} \left|\sum_l e^{ikz_l}\right|^2 \frac{1}{(\varepsilon_\mathbf{k} - \varepsilon_{\mathbf{k-q}}) - \omega_q} \qquad (2.11)$$

If the boson field is the massless field, then $\rho_o = 0$ and Eq. (2.11) becomes

$$E_I = \frac{\hbar c^4}{16R^4} \sum_{q,k} \frac{(\mathbf{qk})^2}{k^4 \omega_q^2} \left|\sum_l e^{ikz_l}\right|^2 \frac{1}{(\varepsilon_\mathbf{k} - \varepsilon_{\mathbf{k-q}}) - \omega_q} \qquad (2.12)$$

Now we apply this equation to a system of two electrons at $\mathbf{z}_1$ and $\mathbf{z}_2$ acting on the vacuum of the massless boson field. In this case

$$\sum_l \left|e^{ikz_l}\right|^2 = 2[1 + \cos(\mathbf{kR})]$$

where $\mathbf{R} = \mathbf{z}_2 - \mathbf{z}_1$, $\varepsilon_\mathbf{k} = \hbar k^2/2m$, $\omega_q = cq$. Neglecting ($\varepsilon_\mathbf{k} - \varepsilon_{\mathbf{k-q}}$) with respect to $\omega_q$, we write

$$\sum_q \frac{(\mathbf{qk})^2}{k^4 \omega_q^3} = \frac{\Omega}{(2\pi)^3} \frac{1}{k^2 c^3} \int_0^\pi \cos^2\alpha \sin\alpha \, d\alpha \int_0^k q \, dq = \frac{R^3}{9\pi c^2}$$

and

$$\sum_\mathbf{k} [1 + \cos(\mathbf{kR})] = 1 + \frac{\Omega}{(2\pi)^2} \int_0^\pi d\theta \int_0^{0.76\pi/R} dk \, k^2 \cos(kR\cos\theta) \sin\theta = 1.514$$

We have considered $\Omega = 4\pi R^3/3$. The upper limit of the ntegral over $k$ appears from the requirement

$$\frac{4\pi R^3/3}{(2\pi)^3} \times \frac{4\pi k^3}{3} = 1$$

The interaction energy becomes [4]

$$E_I = -2 \times \frac{1}{144}\frac{\hbar c}{R} \approx -2\frac{\alpha \hbar c}{R} \qquad (2.13)$$

where $\alpha$ is the fine structure constant. Taking the upper limit of $k$ as $0.94\pi/R$, one obtains the value of $\alpha = 1/137$, jist as for experimental value. The factor 2 appears because we have considered the two neighbours of an electron. Because the interaction energy is attractive, results that the two neighbours have the positive charge. This is an equivalent expression of the Coulomb's law, because $\alpha\hbar c = e^2/4\pi\varepsilon_o$. If in the expression

$$\frac{1}{(\varepsilon_k - \varepsilon_{k-q}) - \omega_q} \qquad (2.14)$$

we replace $\mathbf{q}$ by $-\mathbf{q}$ one obtains

$$\frac{-1}{(\varepsilon_{k+q} - \varepsilon_k) + \omega_q} \qquad (2.15)$$

In the case (2.14) the particles emit bosons, while in the case (2.15) particles absorb bosons. May be shown that the interaction energy may be written in a rather more symmetrical form replacing (2.14) by

$$\frac{\omega_q}{(\varepsilon_k - \varepsilon_{k-q})^2 - \omega_q^2} \qquad (2.16)$$

When a particle has a charge Q, that is Q/e electron charges, we substitute in Eq. (2.11)

$_k = Q_1/e$, $n_{k-q} = Q_2/e$. For $\varepsilon_k - \varepsilon_{k-q} \approx 0$ the sign of the interaction enerqy is minus, that when one particle absorbs and the other emits bosons, the interaction energy is attractive. The expectation value is

$$E_I = -\frac{\alpha \hbar c}{R} \qquad (2.17)$$

where we have considered that interact two particles only. In relations (2.13) and (2.17) do not appear the electric charge, so that we can cosider that this is a general expression for interaction between particles. If these particles are the electrons in the last expression we have the sign plus (the repulsive energy of interaction) and there is the following gauge relation $\alpha\hbar c = e^2/4\pi\varepsilon_o$, where $e$ is the gauge charge, that is the elementary charge of the electron.

If relation (2.17) is applied to gravitation {Newton's law) there is the following gauge relation $\alpha\hbar c = GM^2$, where $G = 6.6726\times10^{-11}$ Nm$^2$kg$^{-2}$ is the gravitational constant and

$$M = \left(\frac{\alpha\hbar c}{G}\right)^{1/2} = 1.859446\times10^{-9}\,kg \qquad (2.18)$$

is the gauge mass. This is a giant particle. If we have a system of the particles with the gauge mass and the gauge charge, this would be be a system of noninteracting particles, a gauge charged superfluid..For two particles with the charges $Q_1$, $Q_2$ and mass $M_1, M_2$, respectibely, the electromagnetic energy of interaction is

$$E(electromagnetic) = \frac{\alpha\hbar c}{r}\frac{Q_1 Q_2}{e^2}$$

and the gravitational energy of interaction is

$$E(gravitational) = -\frac{\alpha\hbar c}{r}\frac{M_1 M_2}{M^2}$$

The ratio

$$\frac{|E(electromagnetic)|}{|E(gravitational)|} = \frac{M^2}{e^2}\frac{Q_1 Q_2}{M_1 M_2} = 1.346933\times10^{20}\frac{Q_1 Q_2}{M_1 M_2} \qquad (2.19)$$

For electrons, this ratio is equal to $4.17\times10^{41}$, a colosal number. The interaction energy (2.17) is attractive. This shows the attractive nature of the interaction in the nature. The force due to gravity, as well as the electromagnetic force between two unlike charges, are always attractive. But the electromagnetic force between two like charges is repulsive. There are positive and negative chrges, while there is positive mass only. This means that the mass is the result of the fluctuations of the energy and (or) of the charge?

## 3. The charged Klein-Gordon field.

The Klein-Gordon equation for free particles is[7]

$$p^\mu p_\mu \Psi = m_o^2 c^2 \Psi \qquad (3.1)$$

where

$$p_\mu = i\hbar \frac{\partial}{\partial x_\mu} = i\hbar \left\{ \frac{\partial}{\partial (ct)}, -\nabla \right\} \qquad (3.2)$$

is the four momentum operator and

$$p^\mu p_\mu = -\hbar^2 \frac{\partial}{\partial x_\mu} \frac{\partial}{\partial x_\mu} = -\hbar^2 \left( \frac{1}{c^2} \frac{\partial^2}{\partial t^2} - \Delta \right) \qquad (3.3)$$

In order to construct the four-currebt $j_\mu$ connected with (3.3) we start from (3.1) with the form

$$(p^\mu p_\mu - m_o^2 c^2) \Psi = 0 \qquad (3.4)$$

and take complex conjugate of this equation, i.e.

$$(p^\mu p_\mu - m_o^2 c^2) \Psi = 0$$

Multiplying both equations from the left, first by $\Psi^*$ and the second by $\Psi$ and calcukating the difference of the resulting two equations yields

$$\Psi^* (p^\mu p_\mu - m_o^2 c^2) \Psi - \Psi (p^\mu p_\mu - m_o^2 c^2) \Psi^* = 0$$

or

$$-\Psi^* (\hbar^2 \nabla_\mu \nabla^\mu + m_o^2 c^2) \Psi + \Psi (\hbar^2 \nabla_\mu \nabla^\mu + m_o^2 c^2) \Psi^* = 0$$

whence

$$\nabla_\mu (\Psi^* \nabla^\mu \Psi - \Psi \nabla^\mu \Psi^*) = \nabla_\mu j^\mu = 0 \qquad (3.1)$$

Therefore, the four-current density is

$$j_\mu = \frac{i\hbar}{m_o} (\Psi^* \nabla_\mu \Psi - \Psi \nabla_\mu \Psi^*)$$

Here we have multiplied by by $i\hbar/m_o$ so that the zero component $j_o$ has the dimensions of a probability density ( that is $1/cm^3$). Eq. (3.1) may be written

$$\frac{\partial}{\partial t}\left[\frac{i\hbar}{m_o c^2}\left(\Psi^*\frac{\partial \Psi}{\partial t} - \Psi\frac{\partial \Psi^*}{\partial t}\right)\right] + div\left(-\frac{i\hbar}{m_o}\right)\left[\Psi^*(\nabla\Psi) - \Psi(\nabla\Psi^*)\right] = 0 \qquad (3.2)$$

By using notations

$$\rho = \frac{i\hbar}{m_o c^2}\left(\Psi^*\frac{\partial \Psi}{\partial t} - \Psi\frac{\partial \Psi^*}{\partial t}\right)$$

$$\mathbf{j} = -\frac{i\hbar}{m_o}\left(\Psi^*(\nabla\Psi) - \Psi(\nabla\Psi^*)\right) \qquad (3.3)$$

eq.(3.2) is the continuity equation

$$\frac{\partial \rho}{\partial t} + div\mathbf{j} = 0$$

$\rho$ is a probability density. However, at a given time $t$, both $\Psi$ and $\partial\Psi/\partial t$ may have arbitrary values and $\rho(x,t)$ may be either positive and negative. Therefore, $\rho(x,t)$ is not positive definite and thus not a probability density. This result is due to the fact that the Klein-Gordon equation is of second order in time , so that we must know both $\Psi(x,t)$ and $\partial\Psi/\partial t$ for a given $t$. The Dirac relativistic wave equation is of first order in time with positive definite probability. We obtain the charge density and the charge current density by multiplication of the expression (3.3) with the elementary charge $e$. The charge in a complex scalar field is

$$Q = \frac{ie\hbar}{m_o c^2}\int d^3x\left(\Psi^*\frac{\partial \Psi}{\partial t} - \Psi\frac{\partial \Psi^*}{\partial t}\right) \qquad (3.4)$$

By interchanging $\Psi$ and $\Psi^*$ we obtain the opposite charge

$$Q = -\frac{ie\hbar}{m_o c^2}\int d^3x\left(\Psi^*\frac{\partial \Psi}{\partial t} - \Psi\frac{\partial \Psi^*}{\partial t}\right)$$

It is observed that in order a charge exist it must that $\Psi$ be a complex function. Relation (3.4) do not define the concept of charge, because we have introduced arbitrarily the elementary charge $e$,

## 4. Generation of the mass by spontaneous breaking of a gauge symmetry

Local U(1) gauge invariance is the requirement that the Lagrangian is invariant under $\Psi' = e^{i\alpha(x)}\Psi$. It is known that this can be achieved by switching the covariant derivation with a special transformation rule for the vector field

$$\partial_\mu \to D_\mu = \partial_\mu - \frac{ie}{\hbar c} A_\mu \quad \text{(covariant derivation)}$$

$$A'_\mu = A_\mu + \frac{1}{e}\partial_\mu \alpha \quad (A_\mu \text{ transformation})$$

(4.1)

The local U(1) gauge invariant Lagrangian for a complex scalar field is then given by

$$L = \frac{\hbar^2}{2m}\left[(D^\mu\Psi)^*(D_\mu\Psi) - \frac{1}{4}F_{\mu\nu}F^{\mu\nu} - V(\Psi^*\Psi)\right] \quad (4.2)$$

The term $(1/4)F_{\mu\nu}F^{\mu\nu}$ is the kinetic term for the gauge field (photon) and $V(\Psi^*\Psi)$ is the extraterm in the Lagrangian

$$V(\Psi^*\Psi) = \frac{m^2c^2}{\hbar^2}(\Psi^*\Psi) + \frac{\lambda c^2}{\hbar^2}(\Psi^*\Psi)^4 \quad (4.3)$$

The Lagrangian (4.1) is studied under small perturbations[8,9]. In an existing theory we are free to introduce a complex scalar field $\Psi = (1/\sqrt{2})(\Psi_1 + i\Psi_2)$ (two degree of freedom). In the case $m^2 > 0$, we have a vacuum at $\begin{pmatrix}0\\0\end{pmatrix}$. The exact symmetry of the Lagrangian is preserved in the vacuum. There are a massless photon and two massive scalar particles $\Psi_1$ and $\Psi_2$ each with a mass $m$. In the case $m^2 < 0$, there is an infinite number of vacua, each satisfying $\Psi_1^2 + \Psi_2^2 = -m^2/\lambda = v^2$. When looking at perturbations around the minimum it is natural to define the shifted fields η and ξ with η = $\Psi_1$ – v, and ξ = $\Psi_2$., which means that the perturbations around the vacuum are described by

$$\Psi_o = \frac{1}{\sqrt{2}}(\eta + v + i\xi) \quad (4.4)$$

Using $\Psi^2 = \Psi^*\Psi = \frac{1}{2}[(v+\eta)^2 + \varsigma^2]$ and $m^2 = -\lambda v^2$ we can rewrite the Lagrangian in

terms of the shifted fields

Kinetic term
$$\frac{2m}{\hbar^2} L_{kin}(\eta,\xi) = \frac{1}{2}(\partial_\mu\eta)^2 + \frac{1}{2}(\partial_\mu\xi)^2 + \frac{ieA_\mu}{2\hbar c}(v+\eta)\partial_\mu\xi + \frac{e^2 A_\mu^2}{2\hbar^2 c^2}v^2 +$$
$$\frac{ie^2 A_\mu^2}{\hbar^2 c^2}v\eta + \frac{e^2 A_\mu^2}{2\hbar^2 c^2}\eta^2 + \frac{ie^2 A_\mu^2}{\hbar^2 c^2}(v+\eta)\xi - \frac{e^2 A_\mu^2}{2\hbar^2 c^2}\xi^2$$

Potential term
$$\frac{\hbar^2}{c^2}V(\eta,\xi) = m^2\Psi^2 + \lambda\Psi^4 = -\frac{1}{4}\lambda v^2 + \lambda v^2\eta^2 + \lambda v\eta^3 +$$
$$\frac{1}{4}\lambda\eta^4 + \frac{1}{4}\lambda\xi^4 + \lambda v\eta\xi^2 + \frac{1}{2}\lambda\eta^2\xi^2$$

The full Lagrangiancan be written as

$$\frac{2m}{\hbar^2}L(\eta,\xi) = \frac{1}{2}(\partial_\mu\eta)^2 - \frac{\lambda c^2}{\hbar^2}v^2\eta^2 + \frac{1}{2}(\partial_\mu\xi)^2 + 0.\xi^2 - \frac{1}{4}F_{\mu\nu}F^{\mu\nu} +$$
$$\frac{1}{2}\frac{e^2}{\hbar^2 c^2}v^2 A_\mu^2 + \frac{evA_\mu}{2\hbar c}(\partial_\mu\xi) + \text{interacting terms}$$

(4.5)

There are the following terms:

1) $\frac{1}{2}(\partial_\mu\eta)^2 - \frac{\lambda c^2}{\hbar^2}v^2\eta^2$     massive scalar particle η

2) $\frac{1}{2}(\partial_\mu\xi)^2 + 0.\xi^2$     massless scalar particle ξ

3) $-\frac{1}{4}F_{\mu\nu}F^{\mu\nu} + \frac{1}{2}\frac{e^2}{\hbar^2 c^2}v^2 A_\mu^2$     massive photon field

However, the Lagrangian also contains strange terms that we cannot easily interpret.- $evA_\mu(\partial_\mu\xi)$. Unlike the η field, describing radial excitations, there is not "force" acting on oscillations along the ξ field. This is a direct consequence of the U(1) symmetry of the Lagrangian and the massless particle ξ is the so-called Goldstone boson.

**5. The unitary gauge assure a generation of a neutral massive particle**

Looking at the terma involving the $\xi$ field (eq. (4.5)) we see that we can rewrite them as

$$\frac{1}{2}(\partial_\mu \xi)^2 - \frac{evA_\mu}{\hbar c}(\partial_\mu \xi) + \frac{1}{2}\frac{e^2 A_\mu^2 v^2}{\hbar^2 c^2} = \frac{1}{2}\frac{e^2 v^2}{\hbar^2 c^2}\left[A_\mu - \frac{1}{ev}(\partial_\mu \xi)\right]^2 = \frac{1}{2}\frac{e^2 v^2}{\hbar^2 c^2}(A'_\mu)^2$$

where we have considered

$$A'_\mu = A_\mu - \frac{1}{e}\left(\partial_\mu \frac{\xi}{v}\right)$$

By comparing this later relation with relation (4.1) results $\alpha = -\xi/v$. This is called the unitary gauge. When choosing this gauge (phase of rotation $\alpha$) the field $\Psi$ changes accordingly

$$\Psi' = e^{-i\xi/v}\Psi = e^{-i\xi/v}\frac{1}{\sqrt{2}}(v + \eta + i\xi) = e^{-i\xi/v}\frac{1}{\sqrt{2}}(v+\eta)e^{i\xi/v} = \frac{1}{\sqrt{2}}(v+f)$$

where we have introduced the real $f$-field. In this case the Lagrangian becomes

$$\frac{2m}{\hbar^2}L = (D^\mu \Psi)^*(D_\mu \Psi) - V(\Psi^* \Psi) = \left(\partial^\mu + \frac{ie}{\hbar c}A^\mu\right)\frac{1}{\sqrt{2}}(v+f)\left(\partial_\mu - \frac{ie}{\hbar c}A_\mu\right)\frac{1}{\sqrt{2}}(v+f) -$$

$$V((v+f)^2) = \frac{1}{2}(\partial_\mu f)^2 - \frac{\lambda c^2}{\hbar^2}v^2 f^2 + \frac{1}{2}\frac{e^2 v^2 A_\mu^2}{\hbar^2 c^2} + \frac{e^2 v A_\mu^2 f}{\hbar^2 c^2} + \frac{1}{2}\frac{e^2 A_\mu^2 f^2}{\hbar^2 c^2} - \frac{\lambda c^2 v f^3}{\hbar^2} - \frac{1}{4}\frac{\lambda c^2 f^4}{\hbar^2}$$

There are the following terms:

1) $\frac{1}{2}(\partial_\mu f)^2 - \frac{\lambda c^2 v^2 f^2}{\hbar^2}$     the massive scalar particle $f$

2) $\frac{1}{2}\frac{e^2 v^2 A_\mu^2}{\hbar^2 c^2}$     the mass term for the gauge field (photon)

3) $\frac{e^2 v A_\mu^2 f}{\hbar^2 c^2}$     photon-Higgs three-point interaction

4) $\frac{1}{2}\frac{e^2 A_\mu^2 f^2}{\hbar^2 c^2}$     photon-Higgs four- point interaction

Because $\partial\Psi/\partial t = \partial(v+f)/\partial t = 0$ as a consequence of Eq. (3.4) results that $Q = 0$. Therefore, the massive particle generated by breaking a local gauge invariant symmetry is electricaly neutral. This particle can decay into charged particles. Also, the perturbation of the unitary gauge can generate charged particles.

**6. The generation of the charged particles by perturbation from the unitary gauge**

If $\alpha \neq \xi/v$ the unitary gauge is not fulfilled and the charged particles can be generated. We have

$$Q = \frac{ie\hbar}{2mc^2}\int d^3x \left[(v+\eta-i\xi)\left(\frac{d\eta}{dt}+i\frac{d\xi}{dt}\right) - (v+\eta+i\xi)\left(\frac{d\eta}{dt}-i\frac{d\xi}{dt}\right)\right] \approx -\frac{e\hbar}{mc^2}\int d^3x\, v\frac{d\xi}{dt}$$

May be written

$$\Psi = Ae^{\beta+i\gamma} \approx A(1+\beta+i\gamma+..) = \frac{1}{\sqrt{2}}(v+\eta+i\xi) = \frac{1}{\sqrt{2}}(v+\eta+\frac{iEt}{\hbar})$$

Therefore,

$$Q = -\frac{e\hbar}{mc^2}\int d^3x\, v\frac{d\xi}{dt} = -\frac{e\hbar}{mc^2}\int d^3x\, v^2\frac{E}{\hbar} = -\frac{eR}{mc^2}\int d^3x\, v^2 = -\frac{emc^2}{mc^2} = -e$$

where we have taken $E = mc^2$. By interchanging $\Psi$ and $\Psi^*$ one obtains the opposite charge. We can, for example, apply these results to the pion triplet ($\pi^+$, $\pi^-$, $\pi^o$). The $\pi^o$ being a neutral particle is characterized by a real wave function

$$\Psi_{\pi^o} = \Psi_o = \frac{1}{\sqrt{2}}(v+f)$$

The $\pi^+$ being charged field, has to be represented by complex wave function

$$\Psi_{\pi^*} = \Psi^* = \frac{1}{\sqrt{2}}(v+\eta-i\xi)$$

The $\pi^-$ has the same mass and oposite charge, so that

$$\Psi_{\pi^-} = \Psi = \frac{1}{\sqrt{2}}(v + \eta + i\xi)$$

**7. Generation of charged particles by decaying of Higgs bosons**

The Higgs boson decays into other particles:
Higgs $\to b + \bar{b}$ ($b$ quark and its antiquark)
Higgs $\to \tau^+ + \tau^-$ ($\tau$ lepton and its antiparticle)
Higgs $\to \gamma + \gamma$ (two photons, also called gamma)
Higgs $\to W^+ + W^-$ (W boson and its antiparticle)
Higgs $\to Z^o + Z^o$ two Z bosons)

The scientists from the CSM experiment with Large Hadron Collider at CERN has succeeded in finding evidence for the direct decay for the Higgs boson into fermions. For example, a Higgs boson decays to two tau particles and further, a taus decays into an electron and a muon. Also, when a proton inside a radionuclide nucleus is converted into a neutron it releases a positron and a neutrino. When a neutron in an atomic nucleus decays into a proton it releases an electron and an antineutrino.

**8. Tesla coil is an user of the excitations of the vacuum**

Tesla coil is used to produce fantastic high voltage long sparking displays [10-11]. It takes the output from 120 VAC to a several kilovolt transformer and driver circuit and steps it up to an extremely high voltage. Voltages can get to be well above 1000 000 volts and are discharged in the form of electrical arcs. Tesla coils are unique in the fact that they create extremeky powerful electrical fields.
Large coils have been known to wirelessly light up fluorescent light up to 50 feet away, and because of the fact that it is an electric field that goes directly into the light and doesn't use the electrodes, even burnt-out fluorescent lights will glow.

**9. Conclusions**

The charge is closely related to the mass. In nature there are some massive charged elementary particles. Composite particles, from positive and negative particles, can be neutral or charged. Because positive and negative charges can neutralize each other, there are composite particles with the same charge and dissimilar masses.

There are two long-range forces in the Universe, electromagnetism and gravity. We have found a general expression for the energy of interaction in these cases: $E_I = \alpha \hbar c / r$, where $\alpha$ is the fine structure constant and $r$ is the distance between the two particles. In the case of the electromagnetic interaction we have $\alpha \hbar c = e^2 / 4\pi\varepsilon_o$, where $e$ is the gauge charge which is the elementary electron charge. In the case of the gravitational interaction we have $\alpha \hbar c = GM^2$ where $M = 1.85 \times 10^{-9}$ kg is the gauge mass of the particle. This is a giant particle. A system of like charged giant particles would be a charged superfluid. All particles with mass attract each other through the exchange of virtual gravitons, which are spin 2 particles. Particles with electric charges can either attract or repel each other by exchanging virtual photons, which are spin 1 particles.

The Higgs massive bosons are generated by spontaneous breaking of a gauge symmetry. The unitary gauge assure the geberation of neutral Higgs particles. The perturbation from the unitary gauges generates the charged massive particles. Also, the Higgs boson decays into charged and neutral particles. Generally, gluons and quark-antiquark pairs, like photons and lepton-antilepton pairs, are excitations of the vacuum.The Tesla coil is an user of the excitations of the vacuum.

**References**
1. F. Halzen and A. Martin, *Quarks and Leptons: An introductory course in Modern Particle Phisics,* Wiley & Sons, 1984
2. C. Quigg, *Gauge theories of the strong, weak and electromagnetic interactions*, Westview Press, 1997
3. D. Griffiths, *Introduction to elementary particles*, Wiley-VCH, 2004
4. Voicu Dolocan, Andrei Dolocan and Voicu Octavian Dolocan, *Qunatum mechanical treatment of the electron-boson interaction viewed as a couling through flux lines*. Plarons, Int. J. Mod. Phys., **B24**, 479-495(2010)
5. Andrei Dolocan, Voicu Octavian Dolocan and Voicu Dolocan, *A new Hamiltonian of interaction for fermions,* Mod. Phys. Lett., **B19**, 6669-681(2005)
6. Ch. Kittel, *Quantum theory of solids*, John Wiley & Sons, 1987
7. W. Greiner, *Relativistic Quantum Mechanics*, Spriger, 2000
8. F. Englert and R. Brout, *Broken symmetries and the mass of gauge vector mesons,* Phys. Rev. Lett., **13**,321-323(1964)
9. P.W. Higgs, *Broken symmetries and the masses of gauge bosons, Phys. Rev. Lett.,* **13**,508-509(1964)
10. Nikola Tesla, *Apparatus for transmission of electrical energy,* US-PatentNo. 645,576, N.Y. 20.3.1900; *Art of transmitting electrical energy through the natural medium.* Us-Patent No.787,412, N.Y. 18.04.1905
11. Konstantin Meyl, *Scalar wave: theory and experiment,* New Hidrogen Technologies and Space Drives, Congres Center Thurgauerhof, CH-8570 Weinfelden, German Association for Space Energy, www.k-meil.de/go/Primaerliteratur/Scalar-waves.pdf